\documentclass[10pt, conference, letterpaper]{IEEEtran} 
\IEEEoverridecommandlockouts
\usepackage[utf8]{inputenc}
\DeclareUnicodeCharacter{202F}{\,}
\usepackage{cite}
\usepackage{amsmath,amssymb,amsfonts}
\usepackage{algorithm}
\usepackage{algpseudocode}
\usepackage{graphicx}
\usepackage{textcomp}
\usepackage{multirow}
\usepackage{booktabs}
\usepackage{makecell} 
\usepackage{longtable}
\usepackage{soul}
\usepackage{xcolor}
\sethlcolor{yellow} 
 \usepackage[colorlinks=true, citecolor=blue, linkcolor=blue, urlcolor=blue]{hyperref}
\usepackage{multirow}
\usepackage{graphicx}
\usepackage{booktabs}
\usepackage{textcomp}
\usepackage{xcolor}
\usepackage{comment}
\usepackage{url}
\usepackage[nobiblatex]{xurl}

\def\BibTeX{{\rm B\kern-.05em{\sc i\kern-.025em b}\kern-.08em
    T\kern-.1667em\lower.7ex\hbox{E}\kern-.125emX}}

\definecolor{warningred}{RGB}{220, 53, 69}       
\definecolor{warningorange}{RGB}{255, 133, 27}    
\definecolor{warningpurple}{RGB}{138, 43, 226}    
\definecolor{warningblue}{RGB}{0, 123, 255}       
\definecolor{warninggreen}{RGB}{70, 161, 4}       

\newif\ifcomment

\ifcomment
\newcommand{\commentmcv}[1]{{\emph{\textcolor{warningred}{[MCV: #1]}}}}
\newcommand{\commentavhi}[1]{{\emph{\textcolor{warningorange}{[Avhi: #1]}}}}
\newcommand{\commentapala}[1]{{\emph{\textcolor{warningpurple}{[Apala: #1]}}}}
\newcommand{\commentee}[1]{{\emph{\textcolor{warninggreen}{[EE: #1]}}}}
\newcommand{\commentnote}[1]{{\emph{\textcolor{warningblue}{[Note: #1]}}}}

\else
\newcommand{\commentmcv}[1]{}
\newcommand{\commentavhi}[1]{}
\newcommand{\commentapala}[1]{}
\newcommand{\commentee}[1]{}
\newcommand{\commentnote}[1]{}

\fi

\newcommand{\bob}{\textsc{ViBe}} 
\newcommand{\bobFull}{VIsion-based BEamforming}

\begin{document}
\bstctlcite{NoDash}

\title{\LARGE Look Once, Beam Twice: Camera-Primed Real-Time Double-Directional mmWave Beam Management for Vehicular Connectivity}



\author{
\IEEEauthorblockN{Avhishek Biswas$^{*,1}$, 
  Apala Pramanik$^{*,1}$, 
  Eylem Ekici$^{2}$, 
  Mehmet C. Vuran$^{1}$}
\IEEEauthorblockA{$^1$\textit{School of Computing,
University of Nebraska--Lincoln}, Lincoln, NE, USA\\
\{abiswas3, apramanik2\}@huskers.unl.edu, mcv@unl.edu}
\IEEEauthorblockA{$^2$\textit{Electrical and Computer Engineering,
The Ohio State University}, Columbus, OH, USA\\
ekici.2@osu.edu}
\thanks{$^*$Avhishek Biswas and Apala Pramanik contributed equally to this work.}
}

\maketitle
\begin{abstract}
Millimeter-wave (mmWave) frequencies promise multi-gigabit connectivity for vehicle-to-everything (V2X) networks, but face challenges in terms of severe path loss and mobility-related beam misalignment. Reliable V2X connectivity requires fast, double-directional beam alignment. However, existing methods suffer from high training overhead and limited generalization to unseen scenarios.
This paper presents \bobFull\ (\bob), a hybrid model-based, closed-loop, learning architecture for real-time double-directional mmWave beam management primed by camera sensing. \bob\ fuses machine learning, model-based reasoning, and closed-loop RF feedback to balance beam-pair establishment latency with link quality. \bob\ bypasses exhaustive training overhead and accelerates link establishment by leveraging camera observations to reduce the beam-search space. Lightweight beam refinement and offset tracking mechanisms adaptively refine beams in response to dynamic application requirements. \bob\ is implemented and evaluated across online indoor/outdoor testbeds, public datasets, and real-time vehicular experiments, demonstrating strong generalization capabilities, making it suitable for real-time V2X communication. Comparisons with 5G NR hierarchical beamforming show that \bob\ consistently maintains lower outage rates. Furthermore, \bob\ outperforms state-of-the-art end-to-end ML models for beam selection when evaluated on public datasets and achieves outage rates as low as $1.1$--$1.4\%$. The results show that a hybrid model-based, closed-loop learning architecture is better suited for real-world mmWave vehicular connectivity than end-to-end trained ML models. For reproducibility, we publish our code to https://github.com/UNL-CPN-Lab/Look-Once-Beam-Twice.

\end{abstract}

\begin{IEEEkeywords}
mmWave, 6G, Beamforming, V2X networks
\end{IEEEkeywords}

\section{Introduction}
\label{sec:intro}

The large-scale deployment of 5G and the evolution toward 6G enable ultra-high data rates, low latency, and ubiquitous connectivity for intelligent transportation systems and V2X networks~\cite{audi5g,audi5gtrack,samsung5g,ntt2025distmimo,chen20235gadvJSAC}.
To support this vision, 3GPP Release~20 emphasizes integrated sensing, beam-based mobility, predictive handovers, and AI-driven decision-making for high-mobility environments~\cite{Lin25Arxiv}.
To this end, millimeter-wave (mmWave) frequencies ($30$--$300$GHz) are promising candidates for next generation vehicular networks, supporting high-rate V2X links~\cite{Tan24COMST}.
mmWave frequencies enable compact antenna arrays at the BS and UE, allowing highly directional beams to overcome severe path loss~\cite{Lockie09MWM}. However, beam misalignment can cause over $10$dB power loss~\cite{Va17TVT}, leading to rapid performance degradation under mobility. Measurements from commercial 5G networks show that only a small subset of available beams is used in practice, with beam refinement largely base station(BS)-centric and inconsistent across operators~\cite{feng2025ACMConxt}. 
\begin{figure}[t!]
    \centering
    \includegraphics[width=0.5\textwidth]{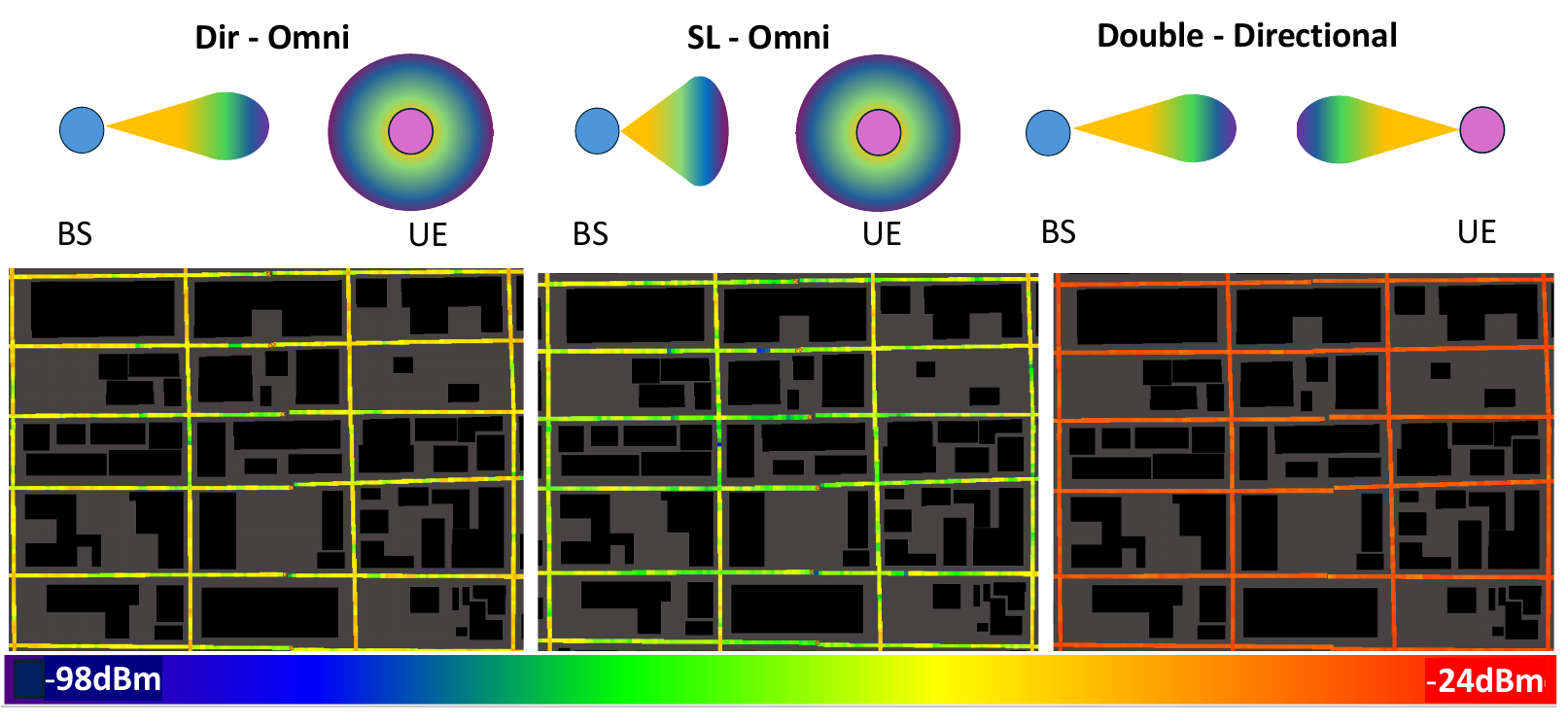}
    \caption{\it Double-directional links improve worst-case received power by 21.8dB and 49.1dB compared to directional-to-omnidirectional (Dir-Omni) and sector-level directional to omnidirectional (SL-Omni) links, improving potential cell size (via Remcom Wireless InSite~\cite{wirelessinsite}). }
    \label{fig:coverage}\vspace{-0.20in}
\end{figure}

Emerging 5G-integrated vehicular platforms increasingly incorporate beam-steerable antenna arrays at both the BS and UE, as shown by recent prototypical and experimental systems~\cite{audi5g,samsung5g}. This enables double-directional beamforming at link access, improving link budget and coverage over single-directional links~\cite{anjinappa2018mmVTC}.
The impact of utilizing double-directional links during the link access stage is illustrated in Fig.~\ref{fig:coverage}, using FR2 urban coverage simulations at 60~GHz in Wireless InSite~\cite{wirelessinsite}. Existing mobile mmWave beam management typically assumes either (a) directional-to-omnidirectional (Dir--Omni) links, where the BS uses directional beams and the UE is omnidirectional, or (b) sector-level directional-to-omnidirectional (SL--Omni) links, where wide BS sector beams (e.g., 5G NR SSBs) reduce acquisition delay. While effective for reducing initial access latency, both approaches \emph{effectively reduce mmWave cell size}. Despite these gains, exhaustive double-directional beam search incurs high latency and computational cost, scaling as $O(N^2)$~\cite{Tan24COMST}, which is prohibitive for V2X applications~\cite{Abari16HotNets}.
Reduced-complexity methods, such as hierarchical, coded, and compressed sensing--based beamforming~\cite{Duan2021KmAccess}, mitigate this overhead but adapt poorly to rapid channel variations, limiting the coverage gains of double-directional access under mobility.

To meet stringent V2X timing and reliability requirements, prior work has focused on reducing beam alignment latency using onboard sensors and end-to-end learning models~\cite{Tan24COMST}; however, these approaches often fail to generalize across environments and mobility conditions. In this paper, we address the problem of low-latency double-directional beam alignment under SNR constraints. We propose \bob: \bobFull, a lightweight and adaptive beamforming framework for vehicular communication.
Unlike end-to-end learning approaches that directly map raw sensory inputs to beam decisions, \bob\ uses a hybrid closed-loop design. The framework combines model-driven beam selection with online SNR feedback.
Initial beam estimation is decoupled from runtime adaptation allowing for a low-overhead coarse beam decision using camera priming and radio coordinate projection. The beam pair is then refined online through a lightweight, iterative SNR-driven feedback.
This design reduces beam management overhead compared to 5G NR and learning-based models. It also improves link reliability. Faster beam alignment and lower outage are achieved under different SNR constraints, therefore not requiring retraining and large-scale labeled RF datasets.

\noindent We validate \bobFull\ through dynamic indoor and outdoor experiments and benchmark \bob\ against current 5G beam management standard~\cite{feng2025ACMConxt} and state-of-the-art methods on public datasets.

\noindent \textbf{Key Contributions.} Our contributions are as follows:
\begin{itemize}
\item We present \bob, a practical, double-directional beam alignment framework that is hardware-agnostic and requires no offline RF training, operating seamlessly with diverse camera setups.

\item \bob\ introduces a hybrid closed-loop adaptation mechanism that combines iterative beam refinement, offset tracking, and a bespoke learning framework, enabling real-time adaptation to SNR dynamics in mobile environments. 

\item We evaluate \bob\ across online indoor/outdoor testbeds, public datasets, and multiple camera configurations, demonstrating impressive generalization capabilities, making it suitable for real-time V2X communication, while significantly reducing beam search overhead.

\item We publicly release the datasets, trained models, and associated code.\footnote{\href{https://github.com/UNL-CPN-Lab/Look-Once-Beam-Twice}{https://github.com/UNL-CPN-Lab/Look-Once-Beam-Twice}} This includes a novel vision model trained on RF street furniture and evaluated in diverse urban scenes.
    
\end{itemize}

\noindent The remainder of the paper is organized as follows. Related work is discussed in Section~\ref{sec:related}, followed by an overview of \bob\ in Section~\ref{sec:overview}. The beam alignment problem and \bob\ design are presented in Sections~\ref{sec:model} and~\ref{sec:ddab}, respectively. Evaluation results and conclusions appear in Sections~\ref{sec:eval} and~\ref{sec:conclusions}.
\section{Related Work}
\label{sec:related}









\noindent Commercial 5G mmWave networks exhibit coverage and performance limits despite dense deployments. In a $2.23$~km$^2$ Chicago area with over $34$ base stations, coverage reached only $35$\%, with throughput degrading under distance and mobility~\cite{Narayanan22INFOCOM} and inefficient spectrum use due to slow inter-band switching~\cite{Liu23INFOCOM}. Beam management further constrains performance by using a few beams (typically $\le5$ of $36$), relying on single-directional gNB-driven refinement, and switching beams infrequently ($0.6$–$2.8$~s)~\cite{feng2025ACMConxt}. Coordinated Tx/Rx steering with narrow beams can instead deliver up to $14$~dB SINR gains and $200$~Mbps higher throughput~\cite{feng2025ACMConxt}, motivating double-directional beamforming.

\begin{figure*}[t!]
    \centering
    \includegraphics[width=\textwidth]{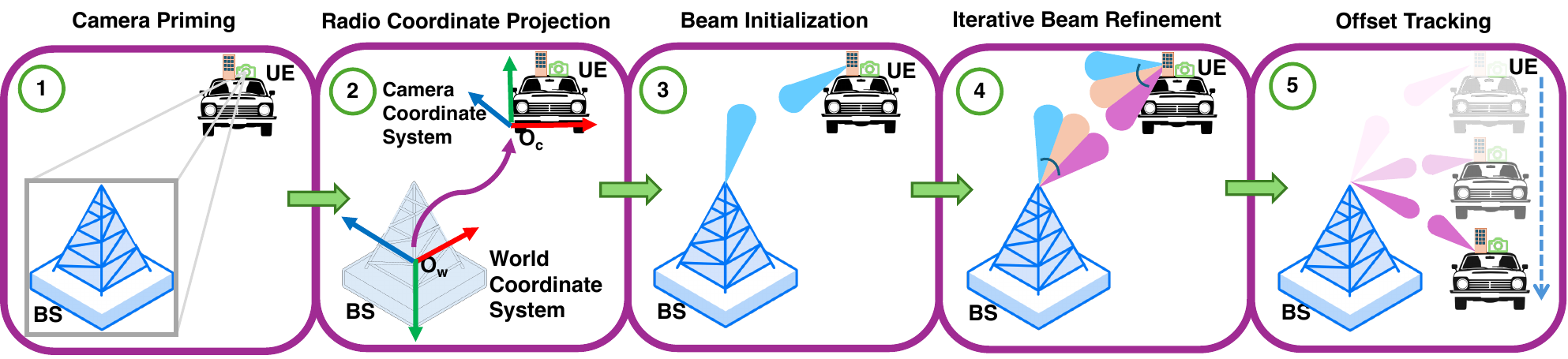}
    \caption{\textit{Overview of} \bob.}
    \vspace{-1em}
    \label{fig:alg_overview}
\end{figure*}
\noindent Double-directional beamforming is computationally expensive due to exhaustive transmit–receive beam pair scanning. Prior work reduces this cost using LBC-based discovery~\cite{shabara2019TON}, Kolmogorov model–based learning~\cite{Duan2021KmAccess}, and compressed sensing or structured search, but these methods rely on static sparsity and degrade under mobility. To mitigate beam alignment overhead under mobility, recent work leverages side information from sensors such as radar, LiDAR, inertial sensors, and cameras. Among these, cameras are particularly attractive due to their low cost and rich spatial information, and have proven effective for line-of-sight beam prediction~\cite{xue24ICS}. However, most vision- and sensor-assisted beam prediction methods remain BS-centric, offline, and non-adaptive, including CNN-based sector prediction using fixed mappings and indoor data~\cite{Salehi20MASS}, image-based beam inference trained on synthetic datasets~\cite{Alrabeiah20VTC}, and learning-driven mobility-aware approaches based on sequence modeling or multi-modal fusion, such as CNN+GRU proactive handover~\cite{Charan21VT}, vision--position fusion for top-\(k\) prediction~\cite{Charan22WCNC}, and LiDAR/GNSS-based recurrent tracking~\cite{Oliveira24LATINCOM}. Moreover, BS-side camera deployment raises privacy concerns~\cite{privacyWashingtonPost}, whereas modern vehicles already integrate multiple cameras for perception and driver assistance~\cite{ADAS}, making UE-side vision a practical and privacy-preserving alternative.

While these deep learning--based methods improve prediction accuracy, they operate as black-box models, lack online RF validation, and are typically evaluated offline, limiting robustness and generalization. Related efforts incorporating inertial sensors, encoder--decoder architectures, semantic awareness, or geometric scene reconstruction further reduce training overhead or input dimensionality~\cite{Zhou17INFOCOM,Jiang22GC,charan2023camera,Imran23ICCWorkshop,Arnold24Arxiv}, but remain largely scenario-specific, and restricted to controlled settings; notably, the semantic-aware approach in~\cite{Imran23ICCWorkshop}, which replaces raw images with object masks or bounding boxes, serves as a key baseline in our evaluation. More recent multicamera and multimodal frameworks improve beam search efficiency and tracking robustness~\cite{Lin24TCom,Zhang22INFOCOM}, while UE-centric methods such as Omni-CNN and FLASH-and-Prune reduce complexity by focusing on SL--Omni links~\cite{Salehi24TVT,Salehi24TMC}, albeit at the cost of reduced cell size and limited adaptability.

\noindent In summary, existing double-directional beamforming approaches reduce overhead at the expense of adaptability or leverage sensing without real-time RF feedback. In Section~\ref{sec:eval}, we evaluate offline end-to-end ML-based methods, which fail to generalize under mobility and dynamic channels, and lack closed-loop adaptation. These gaps motivate a real-time, sensing-driven, and feedback-enabled framework for robust beam alignment in dynamic vehicular environments.

\vspace{-0.4em}
\section{Overview}
\label{sec:overview}
We present an overview of \bob, a camera-primed real-time double-directional link establishment framework for V2X mmWave networks in Fig.~\ref{fig:alg_overview}. \bob\ combines machine learning, model-based reasoning, and closed-loop RF feedback to balance beam alignment latency and link quality. \bob\ distinguishes itself through three innovative approaches:

\textbf{Double-directional Link Establishment.}  
As emerging vehicular prototypes increasingly integrate dedicated 5G/mmWave radios and antenna arrays~\cite{audi5g,samsung5g}, a key challenge emerges: \textit{beam acquisition protocols limit cell size despite the feasibility of double-directional links in V2X networks}. \bob\ addresses this gap with a rapid double-directional link establishment approach that leverages camera-priming to reduce the beam search space.

\textbf{UE-centric Design.} Existing sensor-based beam acquisition solutions primarily assume BS–centric sensors (e.g., cameras, LiDAR, radar)~\cite{Alkhateeb23CommMag,Charan22WCNC,Demirhan22WCNC,Jiang22WCL}, raising privacy concerns that hinder real-world deployment~\cite{privacyWashingtonPost}. 
In contrast, \bob\ adopts a UE–centric design that assumes cameras already deployed in modern vehicles for perception and driver assistance~\cite{ADAS,AutomotiveCameraMarket}. Consequently, \bob\ mitigates privacy concerns, as camera data remains within the vehicle, consistent with current deployment trends. Section~\ref{sec:eval} further shows that \bob\ can be readily deployed in a BS-centric manner while still outperforming existing approaches in delay and performance.

\textbf{Real-time Focus.} \bob\ is designed for real-time operation. Rather than relying on end-to-end machine learning workflows with excessive delays, we introduce a \textbf{hybrid model-based, closed-loop learning architecture} for real-time operation. \bob\ is implemented and evaluated in live vehicle-to-infrastructure experiments. We show that \bob\ can maintain signal-to-noise ratio (SNR) requirements within reasonable delays compared to the state-of-the-art. 

\bob\ consists of five components: \textbf{(1) Camera Priming}, which detects the BS and provides an initial direction estimate to reduce beam search; \textbf{(2) Radio Coordinate Projection}, which maps camera coordinates to radio coordinates using a model-based approach, enabling adaptation across cameras and preserving privacy; \textbf{(3) Beam Initialization}, which converts the direction estimate into UE and BS beam indices compatible with discrete beambooks; \textbf{(4) Iterative Beam Refinement}, which performs a fast local sweep to meet SNR requirements; and \textbf{(5) Offset Tracking}, which maintains residual angle corrections to further reduce beam-pair establishment delay. Experimental results show that this hybrid model-based, closed-loop architecture generalizes effectively under mobility.

\begin{figure}[t!]
    \centering
    \includegraphics[width=\columnwidth]{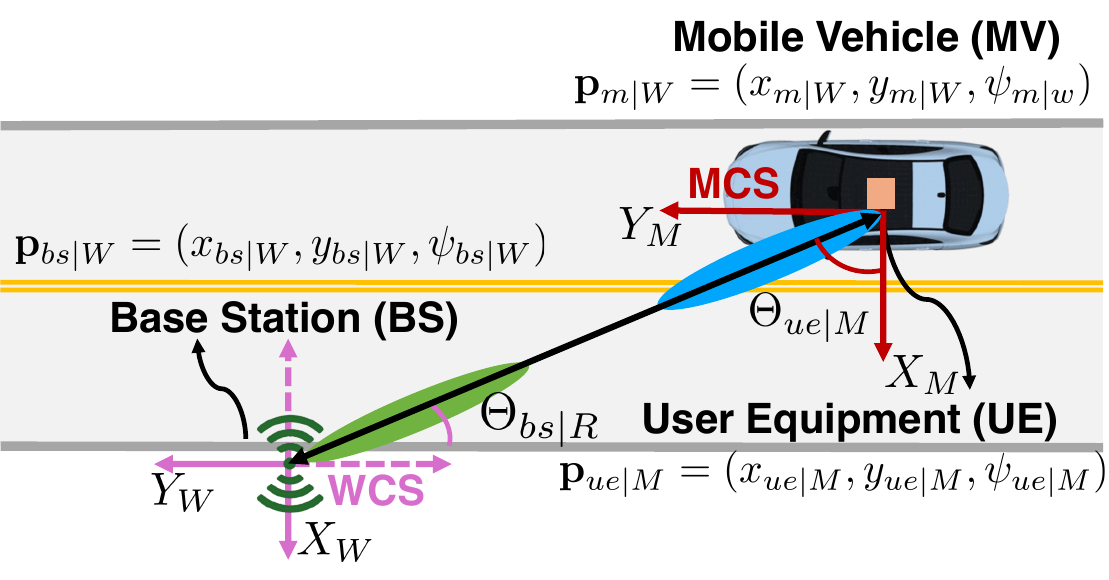}
    \caption{\textit {The V2X communication scenario.}}
    \vspace{-1em}
    \label{fig:system}
\end{figure}
\vspace{-1em}
\section{Preliminaries}
\label{sec:model}
We consider an uplink mmWave V2I communication scenario as shown in Fig.~\ref{fig:system}, where a mobile vehicle (MV), equipped with a radio and a camera, communicates with a base station (BS). We define four coordinate systems that are leveraged throughout the paper: World coordinate system (WCS), mobile coordinate system (MCS), radio coordinate system (RCS), and the camera coordinate system (CCS). 
\vspace{-1em}
\subsection{System Setup}
Within the WCS, the global position of an object (e.g., the mobile vehicle) is defined by its location $\mathbf{p}_{m|W}=[x_{m|W}, y_{m|W}]$, where $X_W$ points East and $Y_W$ points North, and its heading (yaw) angle $\psi_{m|W}$, w.r.t. the $Y_W$ axis. Similarly, the UE, BS, and the camera are defined as $(\mathbf{p}_{ue|W},\psi_{ue|W})$, $(\mathbf{p}_{bs|W},\psi_{bs|W})$, and $(\mathbf{p}_{c|W}, \psi_{c|W})$, respectively. Within the MCS, $(\mathbf{p}_{ue|M},\psi_{ue|M})$ and $(\mathbf{p}_{c|M},\psi_{c|M})$ denote the positions and orientations of the UE and camera, respectively, w.r.t. the vehicle. For both the UE and the BS radios, RCS is used to represent the beamforming angles $\Theta_{ue|R}$ and $\Theta_{bs|R}$ w.r.t. their boresight, respectively. CCS will be utilized to represent the images observed by the camera in the following.


\subsection{Channel Model}
Assuming both BS and the UE are equipped with uniform linear arrays (ULAs) of \(N_B\) and \(N_U\) antennas, respectively, the received signal at the BS is given by~\cite{Alkhateeb14JSTSP}: 
\begin{equation}
    y(t) = \mathbf{w}_{bs}(t)^H \mathbf{H}(t) \mathbf{w}_{ue}(t) s(t) + n(t),
\end{equation}
where \( \mathbf{w}_{ue}(t) \in \mathbb{C}^{N_U \times 1} \) and \( \mathbf{w}_{bs}(t) \in \mathbb{C}^{N_B \times 1} \) are the UE transmit precoder and BS receive combiner, respectively, \( \mathbf{H}(t) \in \mathbb{C}^{N_B \times N_U} \) is the uplink channel matrix, \( s(t) \in \mathbb{C} \) is the transmitted signal, and \( n(t) \sim \mathcal{CN}(0, \sigma^2) \) is the complex Gaussian noise with zero mean and variance \( \sigma^2 \).
Accordingly, SNR is given by:

\begin{equation}
    \gamma(\mathbf{w}_{ue},\mathbf{w}_{bs}) = \frac{\left| \mathbf{w}_{bs}^H(t) \mathbf{H}(t) \mathbf{w}_{ue}(t) \right|^2}{\sigma^2}\;.
    \label{eq:ideal_snr}
\end{equation}

\subsection{Problem Definition}

\textbf{Assumptions.} We assume predefined and fixed beambooks \(\mathcal{B} = \{\mathbf{b}_1, ... ,\mathbf{b}_{|\mathcal{B}|}\}\) and \(\mathcal{U}= \{\mathbf{u}_1, ... ,\mathbf{u}_{|\mathcal{U}|}\}\) at the BS and UE, respectively, with overlapping beams of fixed width. Furthermore, the beambook indices are defined as $k_{bs} \in \{1,...,|\mathcal{B}|\}$ and $k_{ue} \in \{1,...,|\mathcal{U}|\}$, and the beambook beam angles are denoted as $\Theta^{(k_{bs})}_{bs|R}$ and $\Theta^{(k_{ue})}_{ue|R}$. The BS and UE operate under LoS conditions with aligned, parallel boresights. Extension to non-line-of-sight conditions is considered out of scope and constitutes our future work. The boresight assumptions could be easily relaxed through existing pose estimation solutions~\cite{nilsson2014ITS}. The channel matrix \(\mathbf{H}(t)\) is time-varying due to environmental dynamics and mobility, and the optimal transmit/receive beamforming indices \(k_{ue}^*\), \(k_{bs}^*\) are unknown. Additionally, the packet size per beam search is fixed, the SNR constraint, $\gamma_{th}$, is application-specific and given. 

\textbf{Problem.} Accordingly, our goal is to design an online beam-pair selection policy:

\[
\pi: \; (\mathcal{C}, \mathcal{Z}_P, \mathcal{M})\; \longrightarrow (k_{ue}, k_{bs})\;,
\vspace{-1em}
\]
such that

\[
\gamma(\mathcal{U}[k_{ue}], \mathcal{B}[k_{bs}]) \geq \gamma_{th}\;,
\vspace{-0.5em}
\]
where the inputs to the policy are camera observations, $\mathcal{C}$, pilot signal measurements, $\mathcal{Z}_P$, and the internal memory carried over from previous iterations, $\mathcal{M}$. Exhaustive beam sweeping is time-prohibitive in mobile scenarios. Our goal is therefore to use camera observations and accumulated memory to reduce the beam search space while maintaining acceptable link quality under mobility.

\section{\bob : \bobFull}
\label{sec:ddab}

In this section, we present \bob\ (Fig.~\ref{fig:alg_overview}) and describe its five components as illustrated in Fig.~\ref{fig:alg_overview}.


\subsection{Camera Priming}
\label{cam-sense}

\begin{figure}[t!]
     \centering
     \includegraphics[width=0.9\columnwidth]{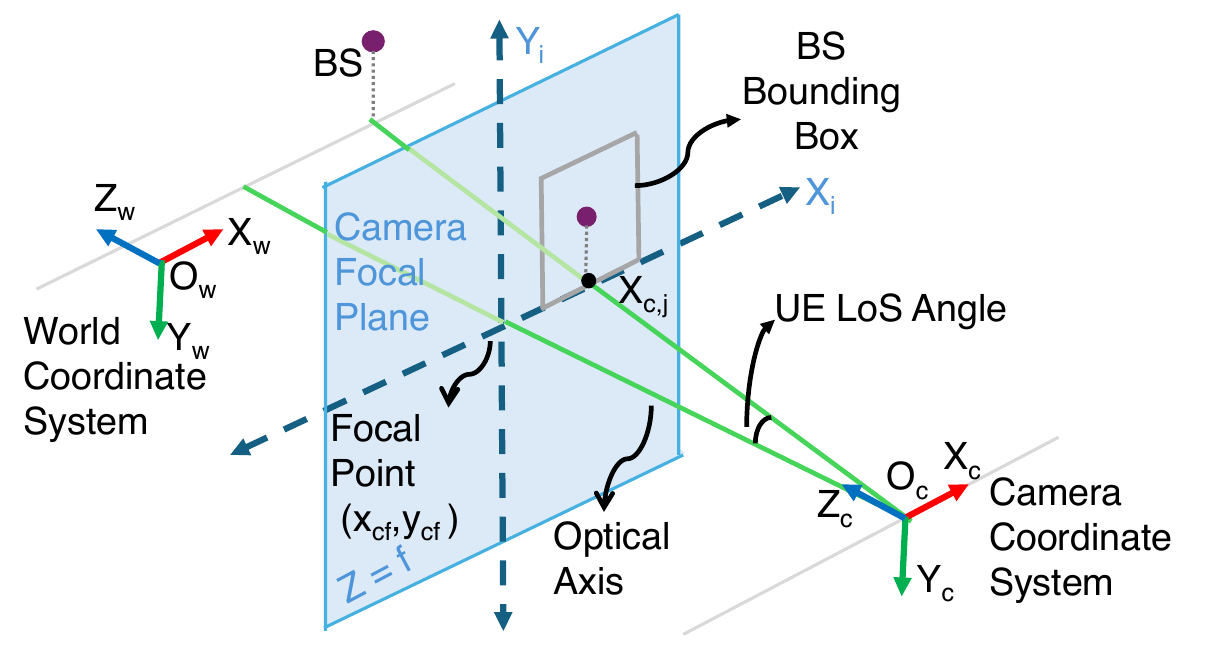}
 \caption{\textit {Pinhole camera model for LoS angle estimation~\cite{Sturm21pinhole}.}} 
 \vspace{-1em}
     \label{fig:PinholeCameraModel}
 \end{figure}

\bob\ reduces the beam-pair search space via camera priming by detecting the BS in the UE camera view. As shown in Fig.~\ref{fig:PinholeCameraModel}, an object detection model trained on street radio furniture processes the images and outputs bounding boxes, \( \mathbf{B} \), as 
\begin{equation}
\mathbf{B} = \left\{(x_{c,j|C}, y_{c,j|C}, \ell_{j}, c_j)\right\}_{j=1}^{J},
\label{eq:boundingbox}
\end{equation}
where \( (x_{c,j|C}, y_{c,j|C}) \) denotes the center pixel coordinates of the \( j \)-th bounding box w.r.t. the camera coordinate system, \( \ell_j \) is the predicted class label (e.g., ``radio''), and \( c_j \in [0, 1] \) is the corresponding confidence score. The total number of detected objects is denoted by \( J \). We assume a single BS is visible in the image, which is reasonable given typical deployment densities. The BS coordinates are then projected into radio coordinates, as described next.

\subsection{Radio Coordinate Projection}  
Upon detection of the BS, the horizontal pixel of the bounding-box center, \( x_{c,j} \), represents the azimuthal displacement of the BS w.r.t. the camera optical axis (Fig.~\ref{fig:PinholeCameraModel}). Accordingly, the estimated LoS angle in the CCS is~\cite{Sturm21pinhole}:

\begin{equation}
    \hat{\theta}_{ue|C}(t) = \operatorname{atan2} \left( \frac{(x_{c,j|C}(t) - x_{cf}) \cdot \mathcal{P}}{f} \right),
    \label{eq:los_from_bbox}
\end{equation}

where $x_{cf}$ is the focal point abscissa, \( \mathcal{P} \) is the pixel pitch (in meters), and \( f \) is the focal length of the camera. Next, we project this estimation first to the MCS and then to the RCS.

The camera and the radio are mounted on the MV with yaws, $\psi_{c|M}$ and $\psi_{r|M}$, respectively. Then, the LoS angle is projected into the RCS as:

\begin{equation}
  \hat{\theta}_{ue|R}(t)=\hat{\theta}_{ue|C}(t)+\psi_{c|M}-\psi_{r|M}.
  \label{eq:theta_rcs}
\end{equation}

This estimation is utilized to initialize the beam pairs. It is important to note that the camera-aided beam estimation is subject to noise from measurement errors, calibration drift, and limited resolution, introducing angular error in the estimated LoS direction. 

\begin{equation}
    \varepsilon_{c}(t) = \theta_{ue|R}(t) - \hat{\theta}_{ue|R}(t), \quad \text{with } |\varepsilon_{c}(t)| \leq \delta_c,
\end{equation}

where \( \delta_c \) denotes the maximum sensor-induced angular deviation under expected operating conditions. Since both the radio and the camera are mounted on the vehicle, the vehicle heading does not affect this projection.

\subsection{Beam Initialization}

The azimuth angle estimate, $\hat{\theta}_{ue|R}(t)$, from the RCS projection is then quantized to find the closest beam index in the UE beambook:

\begin{equation}
    k_{ue}(t) = \arg \min_{k \in \{1:|\mathcal{U}|\}} \left| \Theta^{(k)}_{ue|R} - \hat{\theta}_{ue|R}(t) \right|.
    \label{eq:quantized_beam}
\end{equation}

The UE transmits the index $k_{ue}(t)$ (or equivalently $\Theta^{(k_{ue}(t))}_{ue|R}$) to the BS over a sub-6~GHz control link. Assuming parallel boresights, the BS estimates the beamforming angle at the opposite azimuth:

\begin{equation}
    \hat{\theta}_{bs|R}(t) = \Theta^{(k_{ue}(t))}_{ue|R} + \pi\; ,
    \label{eq:quantized_beam_2}
\end{equation}
and quantizes its prediction similarly. Note that this quantization introduces angular mismatches that are bounded by the half beam spacing of each UE and BS beambooks, as we address next.

\begin{algorithm}[ht!]
\caption{\bob-MA: Beam Selection with Iterative Beam Refinement and Offset Tracking}
\label{alg:bobma_full}
\begin{algorithmic}[1]
\Require Predicted beam $b_{pred}$, SNR Th. $\gamma_{th}$, Guard $\delta$
\If{$b_{pred} \neq \text{None}$}
    \State $\gamma \gets \texttt{MeasureSNR}(b_{pred})$; $n_{beams} \gets 1$
    \If{$\gamma \geq \gamma_{th}$} \Return $(b_{pred}, \gamma, n_{beams})$ \EndIf
    \If{hist[] is not empty}
        \State $b_{adj} \gets b_{pred} + \text{mean(hist[])}$
        \State $\gamma \gets \texttt{MeasureSNR}(b_{adj})$; $n_{beams} \gets 2$
        \If{$\gamma \geq \gamma_{th}$} \Return $(b_{adj}, \gamma, n_{beams})$
        \Else{} \Return \texttt{LocalBeamRefinement}$(b_{adj})$
        \EndIf
    \Else{} \Return \texttt{LocalBeamRefinement}$(b_{pred})$
    \EndIf
\EndIf

\State \textbf{Function} \texttt{LocalBeamRefinement}$(b_c)$
    \State $b^* \gets \text{None}$; $\gamma^* \gets -\infty$; $n_{beams} \gets 0$
    \For{$\delta = 1$ to $\delta_{\max}$}
        \For{$d \in \{-1, +1\}$}
            \State $b \gets b_c + d\delta$; $\gamma \gets \texttt{MeasureSNR}(b)$
            \State $n_{beams} \gets n_{beams} + 1$
            \If{$\gamma \geq \gamma_{th}$} 
            \State Append $d\delta$ to hist[]
            \Return $(b, \gamma, n_{beams})$
            \ElsIf{$\gamma > \gamma^*$} $b^* \gets b$; $\gamma^* \gets \gamma$
            \EndIf
        \EndFor
    \EndFor

\State \Return $(b^*, \gamma^*, n_{beams})$
\end{algorithmic}

\end{algorithm}

\subsection{Iterative Beam Refinement and Offset Tracking}
Camera-primed beam initialization suffers from calibration drift, noise, and beam codebook quantization, while mobility introduces temporal drift that degrades SNR—effects often missed by offline methods. To address this, we designed a fast local beam sweep with refinement and offset tracking, with two variants: \bob-MA, which uses a moving average of past offsets, and \bob-MLP, a lightweight neural network for direct correction.



\textbf{\bob-MA.} The procedure is shown in Algorithm~\ref{alg:bobma_full}. It first checks whether the SNR from beam initialization exceeds the threshold \(\gamma_{\text{th}}\); if so, the predicted beam is accepted. 
If no offset history exists (e.g., at initialization), the UE performs local refinement around the predicted beam using an alternating \(+1, -1, +2, -2, \ldots\) search. For each candidate, the UE signals the beam angle to the BS over a sub-6~GHz link and measures the SNR. The search terminates once a beam meets the threshold, and the resulting offset \(\hat{o}\) relative to the prediction is stored. If no beam satisfies the threshold, the beam with the highest SNR is selected, and the offset history is not updated. The UE maintains an offset history using a moving average of past offsets to correct the predicted beam, yielding \(b_{\text{adj}}\). If \(b_{\text{adj}}\) fails to meet the threshold, the algorithm falls back to local beam refinement.


\textbf{\bob-MLP.} In addition to the rule-based moving average, we evaluate a learned adaptation method, \bob-MLP, which directly predicts the beam offset using a trained black-box model. The network consists of three fully connected layers with LayerNorm, ReLU activations, and dropout, and outputs a single offset value trained using Smooth~L1 loss and the Adam optimizer. At runtime, if the initial SNR falls below the threshold, \bob-MLP infers the corrective offset.
\vspace{-0.6em}

\subsection{Implementation}
\label{sec:implementation}

\begin{figure}[t!]
    \centering
    \includegraphics[width=\columnwidth]{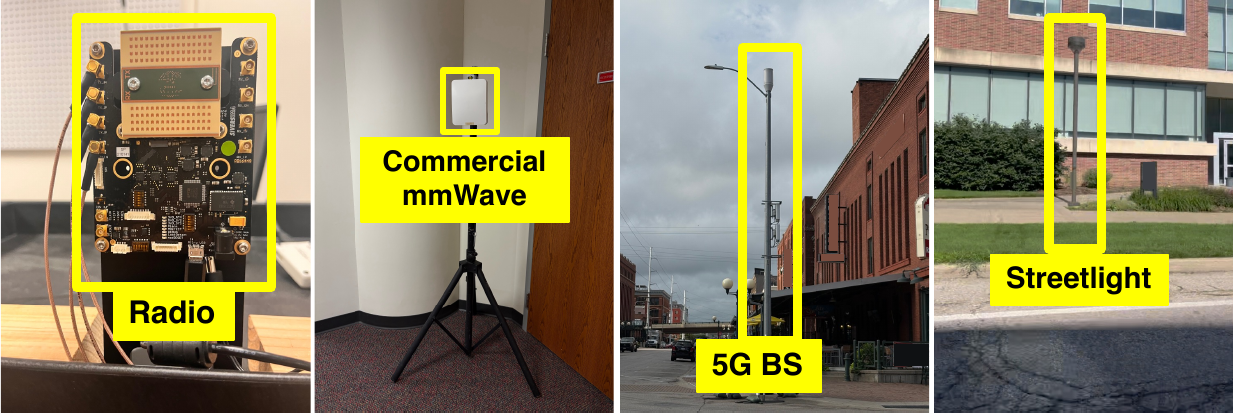}
    \caption{\textit {\bob-YOLOR model sample detections of mmWave base stations and infrastructure in diverse urban scenes at Lincoln, Nebraska,USA.}}
    \vspace{-1em}
    \label{fig:detection}
\end{figure}
To implement the \bob framework, we develop an object
detection pipeline for BS identification for beam initialization using YOLOv11~\cite{khanamyolov11_24archiv}, pre-trained on MS-COCO and fine-tuned on four curated datasets: indoor mmWave radios, commercial mmWave antennas (TG Sounders~\cite{shkel2021configurable}), deployed 5G small cells, and urban streetlights emulating real-world mmWave deployments~\cite{feng2025ACMConxt}. This enables detection of four additional classes beyond COCO, improving adaptability across scenarios (Fig.~\ref{fig:detection}). The resulting detector, combined with the initial beam estimation in Section~\ref{cam-sense}, is referred to as \textbf{\bob-YOLOR} and serves as an internal baseline.

\begin{figure}[t!]
    \centering
    \includegraphics[width=\columnwidth]{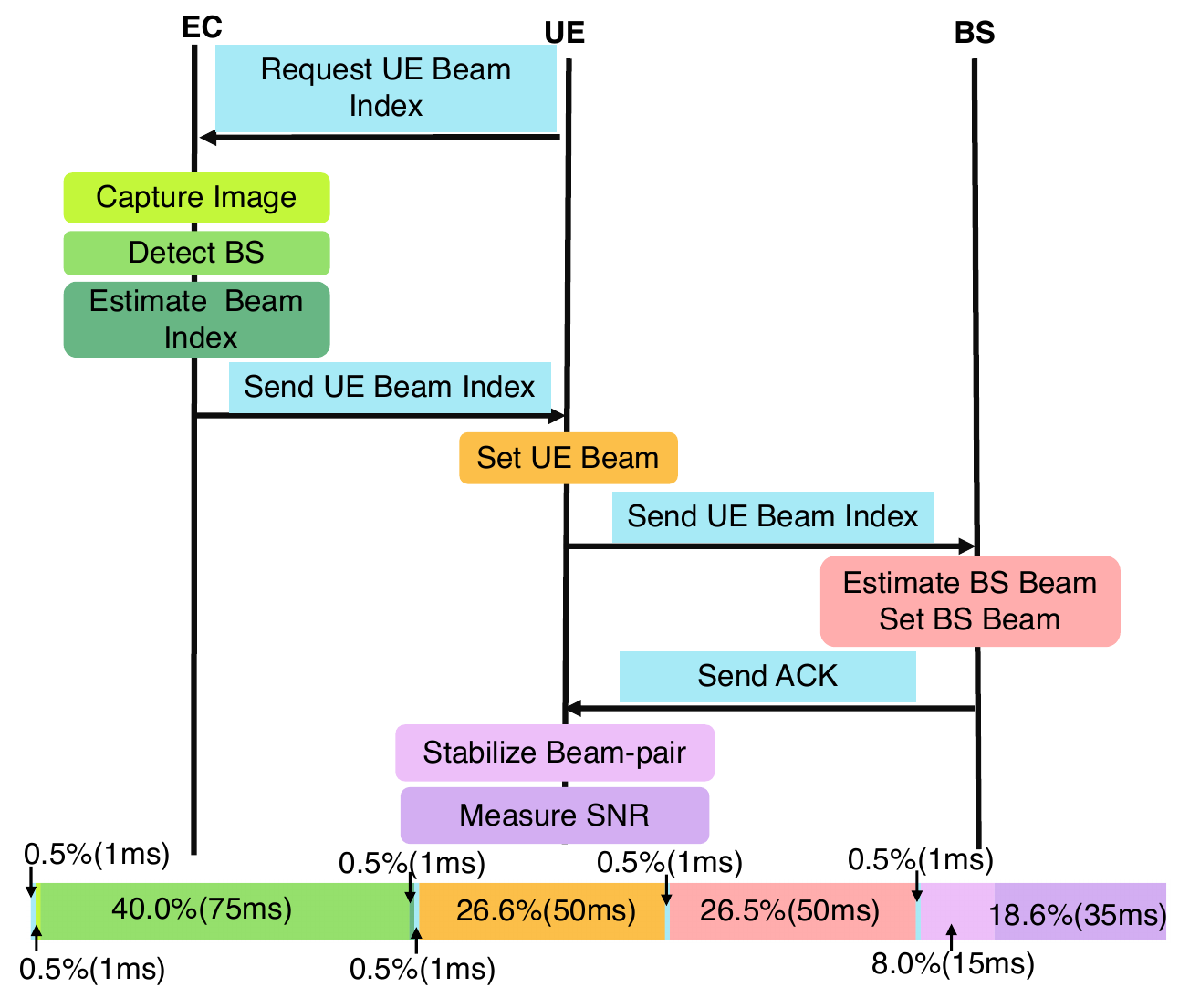}
    \caption{\textit {Timing breakdown of the \bob\ pipeline (Percentages indicate per-step time, EC: Embedded computer at UE).}}
    \vspace{-1em}
    \label{fig:timing}
\end{figure}

Building on this detection capability, we evaluate the runtime efficiency of the \bob\ pipeline. Fig.~\ref{fig:timing} shows an average end-to-end latency of $0.231$s. Image processing accounts for $0.075$s ($40\%$), beam configuration at the UE and BS for $0.050$s ($26.5\%$), and beam stabilization with SNR measurement for another $0.050$s ($26.5\%$). These stages contribute over $90\%$ of the total latency and are primarily hardware dependent, making them key targets for optimization.

\section{Evaluations}
This section presents a comprehensive evaluation of the proposed double-directional beamforming solutions. We first analyze \bob\ and its internal baselines in controlled indoor experiments (Section~\ref{sec:indoor}), then compare \bob\ against state-of-the-art methods on public datasets using outage, coverage, and beam alignment time. Finally, we conduct \textbf{real-time outdoor experiments} to assess latency and outage under dynamic channel conditions (Section~\ref{sec:outdoor}).

\subsection{ Indoor Evaluations}
\label{sec:indoor}
\subsubsection{Experiment Setup}

\begin{figure}[ht!]
    \centering
    \includegraphics[width=1\columnwidth]{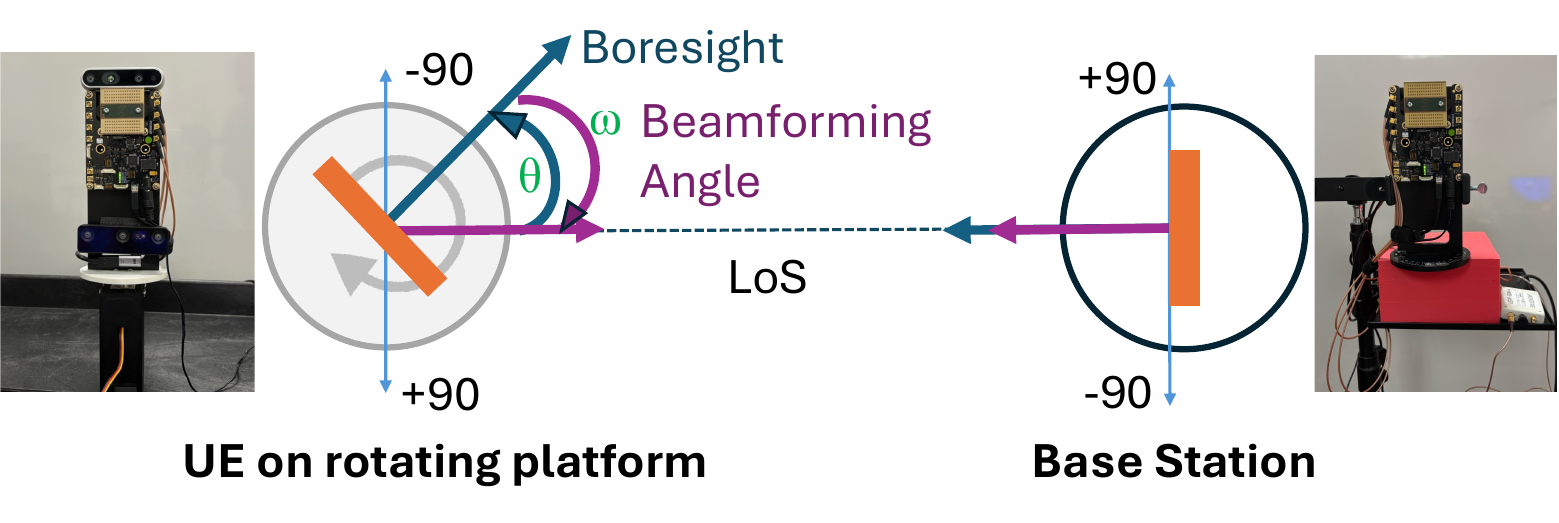}
    \caption{\textit {Indoor testbed at CPN Lab, University of Nebraska-Lincoln, USA.}}
    \vspace{-0.2em}
    \label{fig:internalTopView}
\end{figure}

Indoor experiments are conducted inside the Cyber Physical Networking Lab, Schorr Center, University of Nebraska-Lincoln, using a controlled setup with a fixed BS and a motorized UE to emulate vehicle mobility (Fig.~\ref{fig:internalTopView}). Both nodes employ Sivers Semiconductors 60~GHz EVK06002 phased-array front-ends in the \textit{n263} FR2 band with USRP B200-mini SDRs for baseband processing.

\label{sec:eval}
Each phased-array provides 64 analog beams spanning $\pm 45^\circ$ in $ 1.5^{\circ}$ steps, with half-power beamwidth of $6^\circ$ in azimuth and $18^\circ$ in elevation. The UE is rotated through 180$^\circ$ at angular speeds of $0.25$, $1$, and $4$$^\circ$/s to emulate vehicle motion. Camera-priming is evaluated by either a 60$^\circ$ narrow field of view (NFOV) Intel RealSense camera or a 90$^\circ$ wide FOV (WFOV) Luxonis OAK-D camera.  We evaluate \bob-YOLOR, \bob-MLP, and \bob-MA against ground truth measurements from exhaustive double-directional beam-pair sweeps. Based on the ground truth SNR distributions, three SNR thresholds are defined at the $80^{th}$, $90^{th}$, and $95^{th}$ percentiles, for consistent evaluation.

In the evaluations, \textbf{outage probability} is measured against ground truth, where an outage is recorded if the algorithm fails to select any beam pair exceeding the SNR threshold. The \textbf{beam alignment time}, \(T_b\), is measured using \texttt{clock\_gettime()} from algorithm start until beam selection, at which point the SNR is recorded. Since latency depends on hardware and implementation, the reported delays serve as baseline measurements for fair comparison rather than fundamental limits.
\begin{figure}[ht!]
        \centering
        \includegraphics[width=0.85\columnwidth]{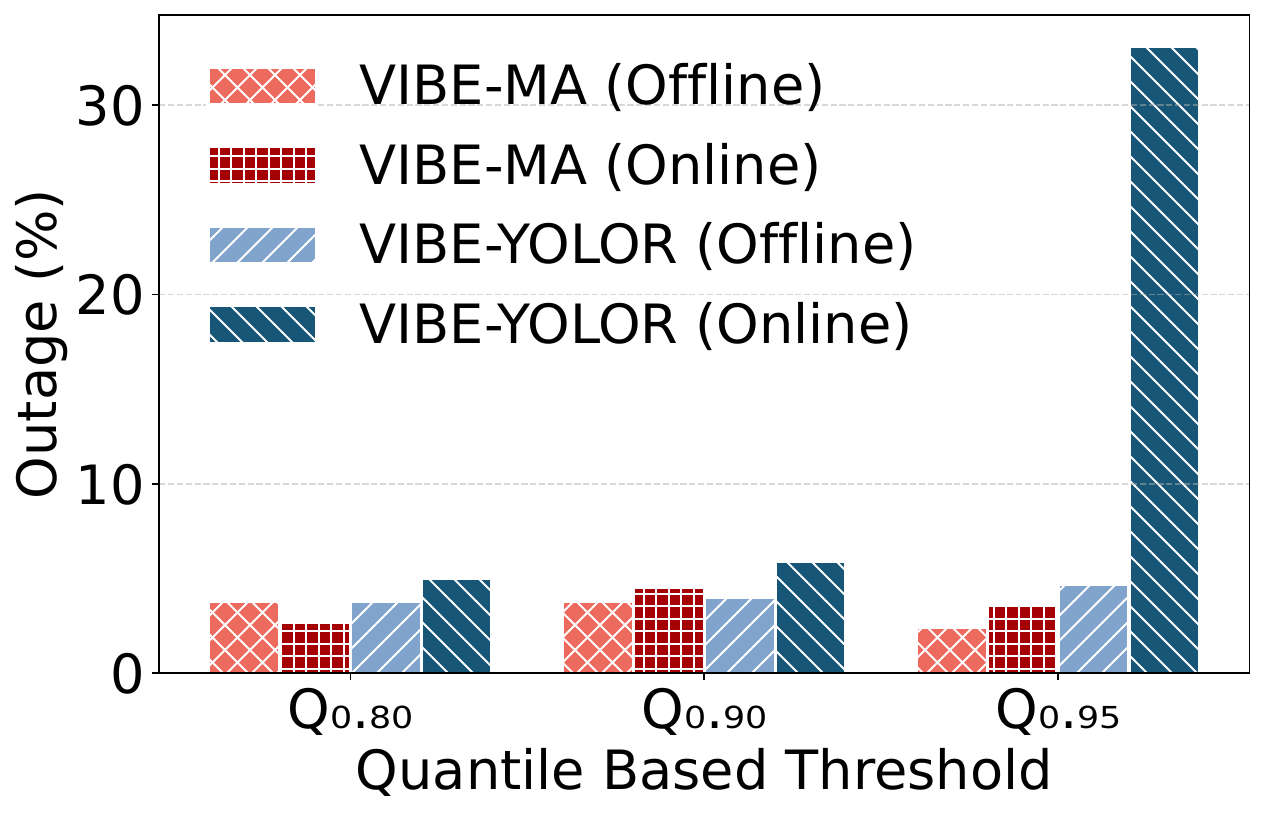}
        \caption{\textit {{Offline vs.~online evaluations.}}}
        \vspace{-0.5em}
        \label{fig:offline-online}
\end{figure}
\subsubsection{Evaluation Results}
\textbf{Offline vs.~Online Evaluations.} 
Recent mobile mmWave studies rely on \textit{offline} evaluations with live images but pre-collected SNR, which omit fast fading and hardware delays. To show this effect, we compare this offline setting with \textit{online} evaluation, where SNR is measured in real time. In Fig.~\ref{fig:offline-online}, we report outage for \bob-YOLOR and \bob-MA across $Q_{0.80}$, $Q_{0.90}$, and $Q_{0.95}$. For \bob-YOLOR, outage at $Q_{0.95}$ increases from $4.7\%$ offline to $33.1\%$ online, showing that offline results overstate reliability. In contrast, \bob-MA outage increases by less than $1.8$~pp across all thresholds. Offset tracking further reduces outage by $0.8$~pp at $0.25^{\circ}/\mathrm{s}$ and $0.5$~pp at $4^{\circ}/\mathrm{s}$, demonstrating robustness under both slow and fast rotations.

\begin{figure}[ht!]
    \centering
    \includegraphics[width=\columnwidth]{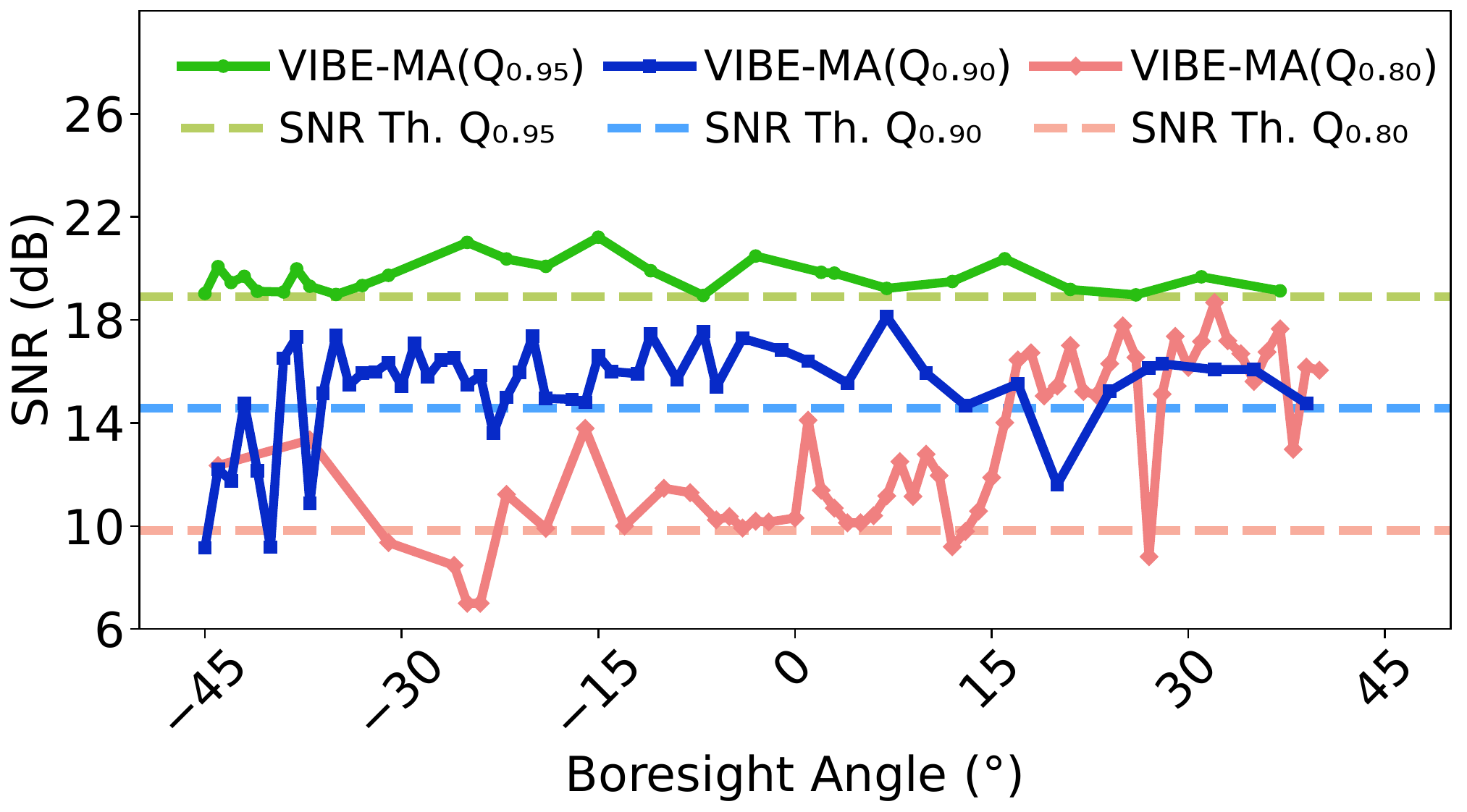}
    \caption{\textit {SNR performance of \bob-MA with WFOV camera at an angular speed of 1$^\circ$/sec.}}
    \vspace{-1em}
    \label{fig:SNR_PLOT}
\end{figure}

\textbf{SNR Adaptation.} In Fig.~\ref{fig:SNR_PLOT}, we present a sample \textit{real-time performance} of \bob-MA with WFOV camera, UE rotating at 1$^\circ$/s, where the dashed lines are the SNR thresholds. It can be observed that, \bob-MA dynamically adapts to the SNR criteria, consistently maintaining higher SNR levels while achieving outages of $13.7$\% ($Q_{0.80}$), $11.7$\% ($Q_{0.90}$), $0$\% ($Q_{0.95}$) on average. This showcases \bob-MA's capability to adapt to different SNR thresholds.

\begin{figure}[ht!]
        \centering
        \includegraphics[width=\columnwidth]{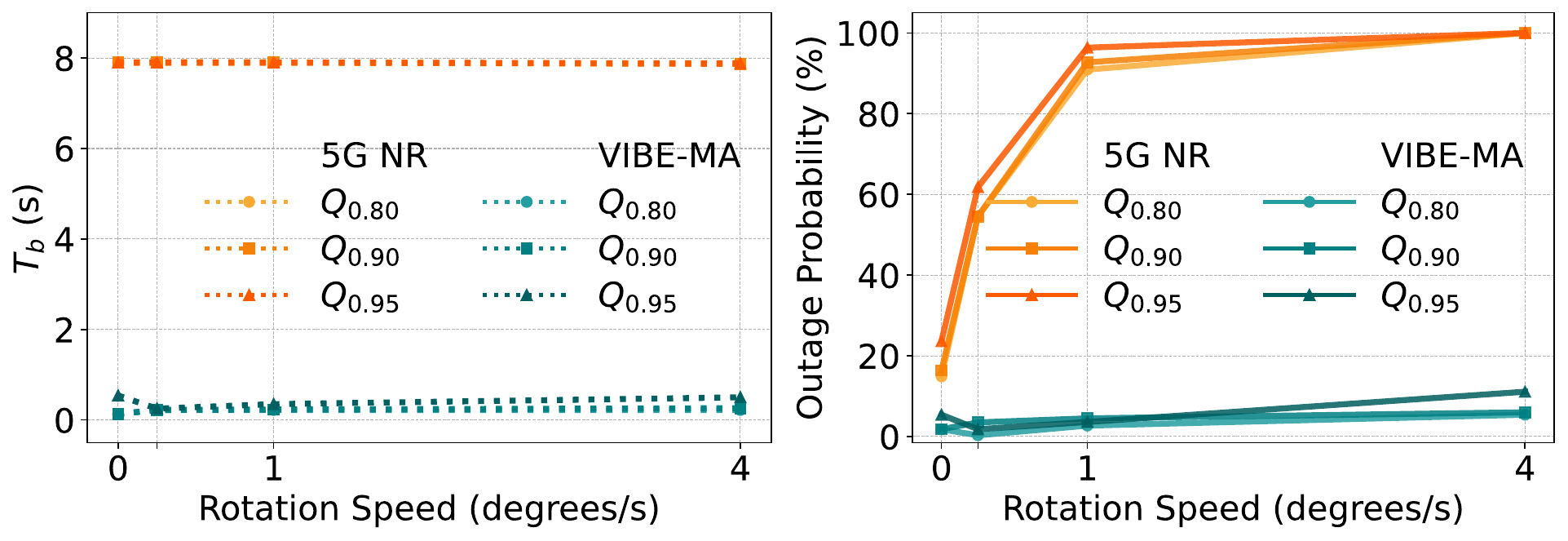}
        \caption{\textit {{Beam alignment time $(T_b)$ and outage probability comparison of \bob-MA with 5G NR}}}
        \label{fig:5g_nr}
\end{figure}

\textbf{Comparison with 5G NR.} 
In Fig.~\ref{fig:5g_nr}, we compare \bob-MA with 5G NR under increasing rotation speeds. 
5G NR shows high SNR outages. Outage probability exceeds \(50\%\) across all quantiles. The beam alignment time remains in the range of \(7.7\!-\!8.0\)~s. 
When beam switching is deferred until the best beam pair is identified (rotation speed = 0), outage reduces to \(14\%\), \(16\%\), and \(23\%\) for \(Q_{0.8}\), \(Q_{0.9}\), and \(Q_{0.95}\), respectively. Outage increases to nearly \(100\%\) as rotation speed increases. 
In contrast, \bob-MA achieves much lower beam alignment time and outage. At the highest rotation speed, \bob-MA maintains \(T_b = 0.5\)~s for \(Q_{0.95}\) with outage remaining below \(12\%\). 
These results show \bob-MA's robustness to mobility-induced angular dynamics. As low-latency beamforming is critical for sustaining connectivity in mobile mmWave scenarios, conventional 5G NR hierarchical beamforming struggles under mobility.

\begin{table}[ht!]
\centering

\caption{\textit {Indoor outage and beam alignment time ($T_{b}$) under varying SNR thresholds and rotation speeds.}}

\renewcommand{\arraystretch}{1.5}
\setlength{\tabcolsep}{4.5pt}
\begin{tabular}{|r|r|
rr|rr|rr||rr|}
\hline
  \multirow{4}{*}{\textbf{SNR}} 
& \multirow{4}{*}{\textbf{Speed}}
& \multicolumn{2}{c|}{\textbf{YOLOR}} 
& \multicolumn{2}{c|}{\textbf{MLP}}
& \multicolumn{4}{c|}{\textbf{MA}}\\
\cline{3-10}
& 
& \multicolumn{2}{c|}{\makecell[c]{NFOV}} 
& \multicolumn{2}{c|}{\makecell[c]{NFOV}} 
& \multicolumn{2}{c||}{\makecell[c]{NFOV}} 
& \multicolumn{2}{c|}{\makecell[c]{WFOV}} 
 \\
Th.
& (deg/s)
& \makecell[c]{Out.\\(\%)} & \makecell[c]{$T_b$\\(s)}
& \makecell[c]{Out.\\(\%)} & \makecell[c]{$T_b$\\(s)}
& \makecell[c]{Out.\\(\%)} & \makecell[c]{$T_b$\\(s)}
& \makecell[c]{Out.\\(\%)} & \makecell[c]{$T_b$\\(s)}\\
\hline
\multirow{3}{*}{$Q_{0.80}$} 
& 0.25 &4.1 &\textit{0.09} &3.8 &0.27 &\textbf{0.3} &0.22 &6.2 &0.13 \\
& 1.00 &5.0 &\textit{0.09} &4.0 &0.22 &\textbf{2.7} &0.22  &13.7 &0.68 \\
& 4.00 &5.6 &\textit{0.09} &5.4 &0.22 &\textbf{5.4} &0.22  &27.2 &1.77 \\
\hline
\multirow{3}{*}{$Q_{0.90}$} 
& 0.25 &4.9 &\textit{0.09} &4.0 &0.22 &3.5 &0.22  &\textbf{2.7} &0.13 \\
& 1.00 &5.9 &\textit{0.09} &\textbf{3.6} &0.23 &4.5 &0.23  &11.7 &1.17 \\
& 4.00 &24.5&\textit{0.09} &7.8 &0.25 &\textbf{6.0} &0.26  &25.0 &1.42 \\
\hline
\multirow{3}{*}{$Q_{0.95}$} 
& 0.25 &22.2 &\textit{0.09} &4.1 &0.26  &\textbf{1.8}  &0.25  &4.4 &0.40 \\
& 1.00 &33.1 &\textit{0.09} &3.4 &0.33  &3.6  &0.35  &\textbf{0.0} &2.72 \\
& 4.00 &64.1 &\textit{0.09} &\textbf{10.0} &0.31 &11.1 &0.50  &28.5 &1.10 \\
\hline
\end{tabular}
\vspace{-2em}
\label{tab:snr-bf-final}

\end{table}
\textbf{Internal Baselines.} Finally, we provide a comprehensive comparison in Table~\ref{tab:snr-bf-final} under different SNR thresholds and speeds, and different cameras in an online setting. In majority of the cases, \bob-MA achieves the lowest outage, reducing outage by up to $29.5$pp (e.g., from $33.1\%$ to $3.6\%$ at $Q_{0.95}$ and $1^\circ$/s) albeit with an increase in alignment time from $0.09$s to $0.35$s. \bob-YOLOR maintains a constant beam alignment time, which may be desirable in low-speed and low SNR threshold conditions. While \bob-MLP occasionally outperforms \bob-MA (e.g., $3.6\%$ outage at $Q_{0.90}$, $1^\circ$/s), the differences are marginal and inconsistent, emphasizing \bob-MA's overall generalizability.
Compared to \bob-MLP, which is trained on prior indoor offset data, \bob-MA reduces outage by up to $1.8$~pp while achieving similar beam alignment times, highlighting the benefit of its hybrid closed-loop design for meeting real-time SNR thresholds.
Furthermore, \bob-MA is \textit{hardware agnostic}. When tested with a WFOV camera, \bob-MA achieves $0\%$ outage at $Q_{0.95}$, $1^\circ$/s, confirming its robustness across different sensing configurations. 
At higher speeds ($4^\circ$/s), the WFOV configuration incurs a $17.4$~pp higher outage than NFOV ($11.1\%$) due to reduced angular resolution. Coarser quantization under fast motion increases beam uncertainty and corrective search time, raising \(T_b\) from $0.50$~s (NFOV) to $1.10$~s (WFOV).
Overall, \bob-MA reduces outage by up to $53$pp and $3.5$pp compared to \bob-YOLOR and \bob-MLP, respectively. Although this incurs a modest increase in alignment time, it remains suitable for real-time operation (Section~\ref{sec:outdoor}). The closed-loop hybrid design is hardware agnostic and robust in real-time scenarios.

\subsection{State-of-the-art Comparisons}
\label{sec:comparison}
\subsubsection{Experiment Setup}
 To evaluate robustness and cross-scenario generalization, we compare \bob-YOLOR and \bob-MA against two state-of-the-art baselines: MobileNet+LeNet (MNet–LeNet)~\cite{Imran23ICCWorkshop}, trained on Scenario~7 of~\cite{Alkhateeb23CommMag}, and ResNet-50~\cite{Charan22WCNC}, trained on Scenario~6 of~\cite{Alkhateeb23CommMag}. Both baselines are evaluated using standard top-$k$ beam prediction accuracy (e.g., top-1 and top-3), with generalization tested on unseen Scenario~9~\cite{Alkhateeb23CommMag}. To ensure a fair comparison, we also evaluate \bob\ in BS-centric configurations, demonstrating its adaptability beyond UE-centric operation.
It is important to note that YOLOR was \textit{not} trained in any of these scenarios, making all of them \textit{unseen}. Model performance is evaluated in terms of outage, which is based on the number of instances where the predicted beam power falls under the received power threshold, and beam alignment time, which accounts for image inference and beamforming delay derived from indoor measurements. We deploy the open-source models, as is, locally and all evaluations are conducted on an NVIDIA A2000 GPU (12GB VRAM) using quantile-based \textit{normalized received power} threshold of ($Q_{0.80}$, $Q_{0.90}$, and $Q_{0.95}$) because SNR information was unavailable in the datasets. 

\subsubsection{Evaluation Results}
The results for \bob-MA, MNet-LeNet, and ResNet-50 across both seen and unseen scenarios are shown in Table~\ref{tab:ExternalEval_LeNet_Resnet}. The first three rows report \bob-MA results averaged across Scenarios 6 and 7, on which \bob-MA is not trained. In the bottom rows, we report performance on Scenario~9, which remains unseen for all models.
When evaluated on the scenarios of baseline models, \bob-MA consistently achieves low outage with an average beam alignment time below $0.26$s. Unlike the baselines, \bob-MA maintains stable performance across thresholds and scenarios. Despite not being trained on the same datasets, \bob-MA maintains a \textbf{very low outage of 1.3\%-1.4\%}. On Scenario~7, \bob-MA achieves $24.7$pp to $57.1$pp lower outage than MNet-LeNet as threshold increases from $Q_{0.80}$ to $Q_{0.95}$. On Scenario~6, \bob-MA performs comparably to ResNet-50, which achieves $0\%$ Top-3 outage at lower thresholds. However, at $Q_{0.95}$, ResNet-50 outage increases to $1.5\%$, while \bob-MA remains at $1.3\%$, \textbf{with 143\% faster beam alignment}.
\begin{table}[t!]
\centering
\caption{\textit 
{Outage probability and beam alignment time for \bob-MA against MNet+LeNet~\cite{Imran23ICCWorkshop} (trained on Scen.~7) and ResNet-50~\cite{Charan22WCNC} (trained on Scen.~6). Scen.~9 serves as an unseen environment for all models.}}
\label{tab:ExternalEval_LeNet_Resnet}
\renewcommand{\arraystretch}{1.9}
\setlength{\tabcolsep}{1.5pt}
\resizebox{\columnwidth}{!}{%
\begin{tabular}{|c|c|c|c|rrr|c|rrr|r|}
\hline
\multirow{3}{*}{\rotatebox{90}{\textbf{Scenario}}}&
\multirow{3}{*}{\textbf{\begin{tabular}[c]{@{}c@{}}Norm.\\ $P_r$ Th.\end{tabular}}} &
\multicolumn{2}{c|}{\textbf{\bob-MA}} &
\multicolumn{4}{c|}{\textbf{MNet + LeNet~\cite{Imran23ICCWorkshop}}} &
\multicolumn{4}{c|}{\textbf{ResNet-50~\cite{Charan22WCNC}}} \\
\cline{3-12}
& & \textbf{Out.} & \textbf{$T_{b}$} & \textbf{Top-1} & \textbf{Top-2} & \textbf{Top-3} & \textbf{$T_{b}$} 
& \textbf{Top-1} & \textbf{Top-2} & \textbf{Top-3} & \textbf{$T_{b}$} \\
& & (\%) & (s) & (\%) & (\%) & (\%) & (s) & (\%) & (\%) & (\%) & (s) \\
\hline

\multirow{3}{*}{\rotatebox{90}{\textbf{6,7}} }
& $Q_{0.80}$ & 1.4 & 0.23 & 44.1 & 33.6 & 26.1 & \textit{0.17} & 0.8  & 0.1 & \textbf{0.0} & 0.56 \\
& $Q_{0.90}$ & 1.3 & 0.23 & 57.0 & 52.1 & 47.2 & \textit{0.17} & 5.7  & 0.3 & \textbf{0.0} & 0.56 \\
& $Q_{0.95}$ & \textbf{1.3} & 0.26 & 73.5 & 63.6 & 58.4 & \textit{0.17} & 53.8 & 15.8 & 1.5 & 0.56 \\
\hline

\multirow{3}{*}{\rotatebox{90}{\textbf{9}} }
& $Q_{0.80}$ & \textbf{1.0} & 0.20 & 46.7 & 37.1 & 32.1 & \textit{0.17} & 79.8 & 63.4 & 50.5 & 0.40 \\
& $Q_{0.90}$ & \textbf{1.1} & 0.23 & 70.5 & 55.8 & 49.1 & \textit{0.17} & 91.1 & 83.4 & 75.6 & 0.41 \\
& $Q_{0.95}$ & \textbf{1.1} & 0.23 & 84.1 & 70.6 & 65.3 & \textit{0.17} & 95.1 & 91.1 & 86.6 & 0.40 \\
\hline
\end{tabular}
} 
\vspace{-2em}
\end{table}

\textbf{Generalization.} In the unseen Scenario 9, \textbf{both baselines struggle to generalize:} MNet-LeNet Top-3 outage exceeds $65\%$, and ResNet-50 leads to $85.6\%$ outage. On the other hand, \bob-MA sustains outage of only $1.1\%$, with up to $69.9$pp lower outage and $79\%$ lower latency than ResNet-50. These results show that \bob-MA can generalize across environments and thresholds, while highlighting the limitations of black-box ML models trained on specific scenarios.

\begin{figure}[t!]
    \centering
    \includegraphics[width=\columnwidth]{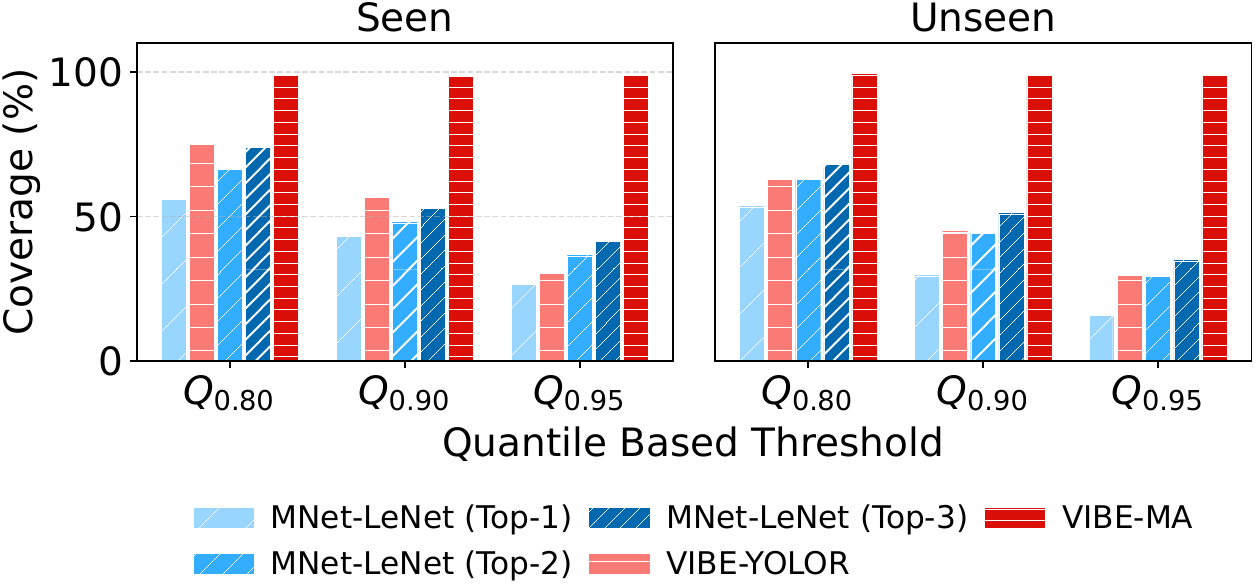}
    \caption{\textit {Coverage percentage [100-(Outage \%)] across thresholds in seen and unseen scenarios.}}
    \vspace{-1em}
    \label{fig:MNET-LENET-YOLOR_SuccessRate}
\end{figure}

In Fig.~\ref{fig:MNET-LENET-YOLOR_SuccessRate}, we compare the coverage percentage [$100-(\text{Outage}~\% )$] of \bob-MA and \bob-YOLOR with the Top-K predictions from MNet-LeNet under varying thresholds. \textbf{\bob-MA consistently results in the highest coverage (98.6\%-98.9\%)}, in all thresholds and even under challenging conditions that are not within its training set. Compared to the strongest baseline, MNet-LeNet Top-3, \bob-MA reduces outage by up to $69.3$pp. The coverage gains of \bob-MA are partly attributed to \bob-YOLOR, which encapsulates the first three stages of \bob-MA. \bob-YOLOR outperforms MNet–LeNet Top-1 by $9.5$–$13.7$~pp and achieves coverage comparable to Top-2 using a single beam decision. Although Top-3 attains up to $11.5$~pp lower outage than \bob-YOLOR, it requires an additional beam selection step and still underperforms \bob-MA.

The generalization gap can be observed in Fig.~\ref{fig:MNET-LENET-YOLOR_SuccessRate} when MNet-LeNet outage is compared for seen and unseen scenarios, where \bob-MA is virtually unaffected. Furthermore, the coverage of \bob-MA is not affected as the threshold increases, where other methods suffer at higher quantiles. The results highlight the importance of closed-loop feedback and the generalization ability of \bob-MA.
Overall, \bob-MA achieves a strong balance between low outage and low latency in BS-centric settings, despite not being trained on any evaluation scenarios. This highlights the suitability of hybrid model-based, closed-loop architectures over end-to-end ML for real-world mmWave vehicular connectivity.

\subsection{ Outdoor Evaluations}
\label{sec:outdoor}

\subsubsection{Experiment Setup}
\begin{figure}[ht!]
    \centering
\includegraphics[width=1\columnwidth]{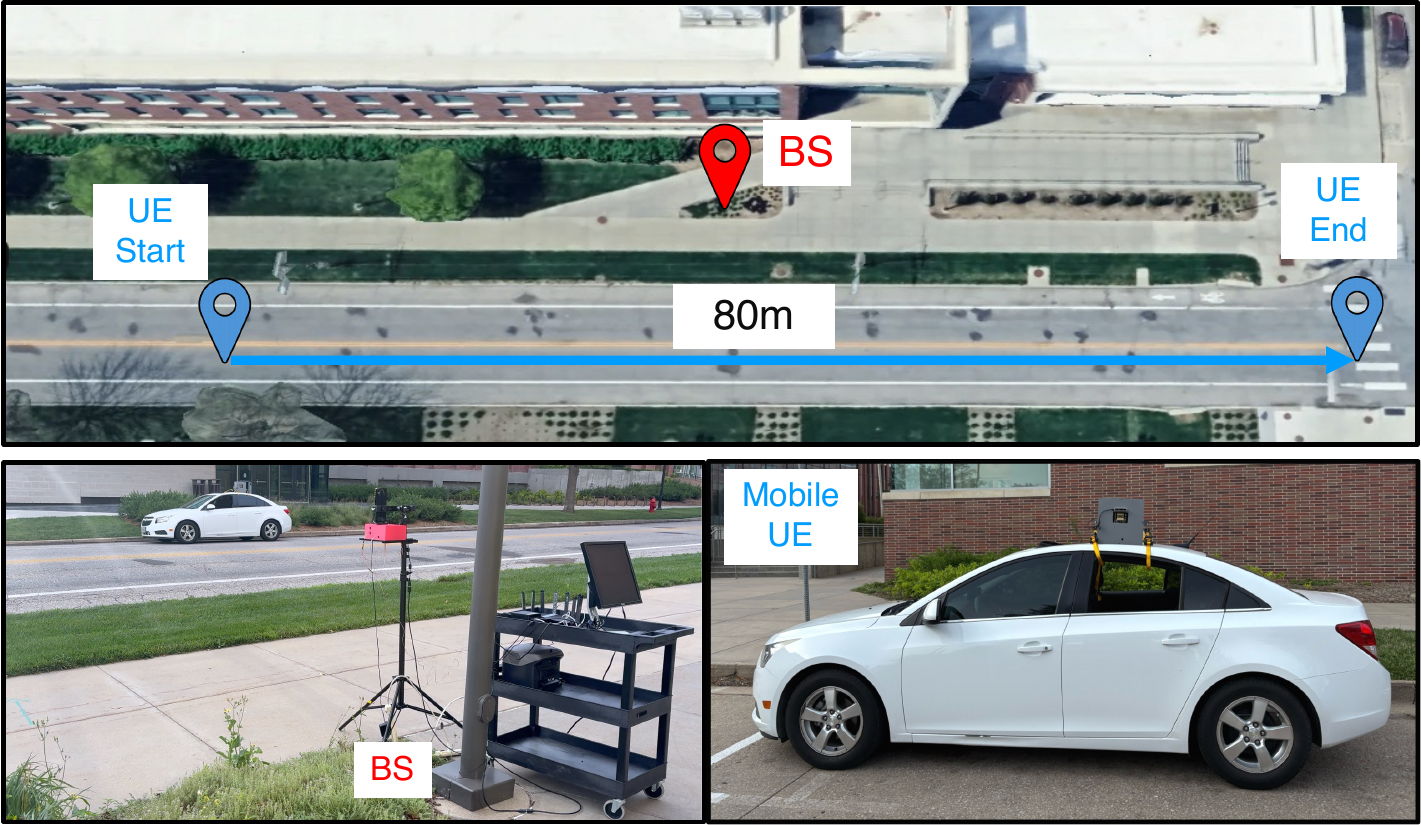}
    \caption{\textit {Outdoor Evaluation Testbed at University of Nebraska-Lincoln, USA.}}
    \vspace{-1em}
    \label{fig:outdoorsetup}
\end{figure}

The outdoor experiment is conducted on the University of Nebraska-Lincoln campus, using a fixed BS and a mobile UE moving along an 80-m straight path, as shown in Fig.~\ref{fig:outdoorsetup} (top). To emulate a worst-case V2I scenario, the UE boresight is oriented perpendicular to the road, causing rapid beam angle variations during motion. The UE detects urban streetlights—representing mmWave BS deployments in U.S. cities~\cite{feng2025ACMConxt}—using a $60^\circ$ NFOV Intel RealSense camera, and performs real-time beam selection and refinement. Experiments are conducted for \bob-YOLOR, \bob-MA, and \bob-MLP at angular velocities of $1.6^\circ$/s, $8.0^\circ$/s, and $12.8^\circ$/s.
\subsubsection{Evaluation Results}
\begin{figure}[t!]
    \centering
\includegraphics[width=\linewidth]{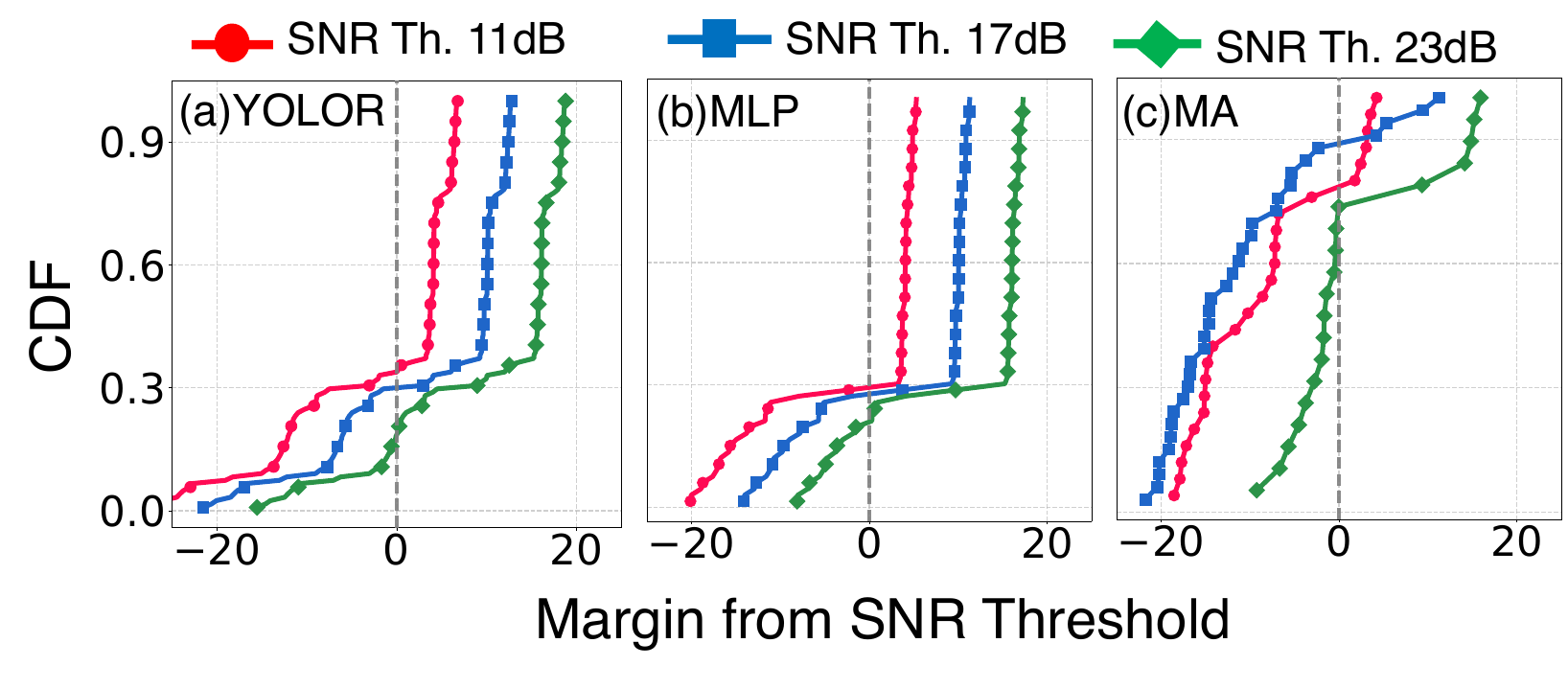} 
    \caption{\textit {Outdoor evaluations: CDF of margin from SNR threshold ($x=0$): (a) \bob-YOLOR, (b) \bob-MLP, and (c) \bob-MA (8.0$^\circ$/s or $5$mph).}}
    \vspace{-1em}
    \label{fig:outdooreval}
\end{figure}

In Figs. \ref{fig:outdooreval}, we show the CDFs of the margin from SNR threshold ($11$dB, $17$dB, and $23$dB). More specifically, we plot the CDF of $\gamma_{th} - \gamma$ (i.e., $F(\gamma_{th} - \gamma) = P(\gamma_{th} - \gamma \leq x)$), which essentially shows the probability that the achieved SNR is above or equal to the SNR threshold, while normalizing different SNR thresholds to $x=0$ on the plot. When the vehicle is traveling at the angular velocity of 8.0$^\circ$/s, \bob-YOLOR meets the SNR threshold in only $33.9$\%, $30.0$\%, and $18.5$\% of cases for SNR thresholds of $11$dB, $17$dB, and $23$dB, respectively, showing a steep decline as the SNR threshold increases. \bob-MLP performs slightly better for the highest SNR threshold but worse for others as compared to \bob-YOLOR, achieving $29.4$\%, $27.8$\%, and $21.0$\%. In contrast, \textbf{\bob-MA achieves substantially higher real-time reliability}, meeting the SNR threshold in $ 78.5$\%, $89$\%, and $73.7$\% of cases. This performance can further be enhanced through advanced techniques such as adaptive modulation and coding, offering the potential for full connectivity in mmWave V2X networks.

\textbf{Practical Considerations.}
Some performance metrics, including latency and coverage, are hardware-dependent and can be improved through further engineering, which is beyond the scope of this paper. Specifically, the end-to-end latency of \bob\ is constrained by the inference speed of the vision pipeline and the beam switching rate of the phased-array hardware, which together bound the maximum vehicular speed the system can support. As a result, \bob\ is evaluated at angular velocities corresponding to vehicle speeds of 1, 5, and 8~mph—representative of low-speed urban and parking scenarios. These are not fundamental limitations of the \bob\ framework, but rather artifacts of the current prototype hardware.

Scalability to higher speeds is achievable along two independent axes. First, extending camera range increases the distance at which a BS becomes visible, which reduces the rate of angular change experienced during approach and thereby relaxes latency requirements. In our current setup, a camera with a 3~m focal length enables BS detection up to 16~m; replacing this with existing ADAS-grade cameras that support ranges up to 100~m \cite{otobrite} would enable operation at speeds of approximately 50~mph. Second, inference latency can be independently reduced by deploying on more capable edge platforms: YOLOv11x latency drops from 75ms on a Jetson Orin Nano to 20~ms on an NVIDIA DRIVE AGX \cite{nvidia}, directly translating to higher supportable speeds. Together, these improvements suggest a clear and practical path toward highway-speed deployment using commercially available hardware.

\textbf{Summary.} \bob-MA, a \textit{hybrid model-based, closed-loop learning architecture}, consistently outperforms competing methods. Indoors, \bob-MA reduces outage by up to $53$pp and $3.5$pp over internal baselines. Against state-of-the-art baseline methods on unseen datasets, \bob-MA outperforms MNet-LeNet by achieving $47.7$pp lower outage. Compared to ResNet-50, \bob-MA has $69.9$pp lower outage and is $73.9\%$ faster. Finally, in outdoor, real-time trials, \bob-MA lowers outage by up to $59$pp compared to its internal baselines. These results establish \bob-MA as a \textit{robust, scene-agnostic solution} for a reliable beam alignment.

\section{Conclusions}
\label{sec:conclusions}

We present \bob, which combines visual sensing with lightweight online correction to enable fast beam acquisition and reliable link maintenance without large-scale RF training data. Extensive indoor, outdoor, and cross-scenario evaluations show consistently low outage and strong robustness in unseen environments, where 5G NR hierarchical beamforming and black-box ML models struggle. \bob\ provides a practical, low-cost alternative to radar, LiDAR, and GNSS using camera priming.

\textbf{Limitations and Future Work.} We plan to further improve \bob\ timing performance through hardware and software optimizations for high-velocity operation. The UE–BS coordination can be extended to arbitrary BS alignments using pose estimation. While this work focuses on line-of-sight scenarios, \bob\ can be extended to non-line-of-sight settings using historical beam measurements and structural awareness. Camera impairments in high-mobility environments can be mitigated through multi-camera configurations or predictive tracking.
\textbf{Impact.} Beyond performance gains, \bob\ confines visual sensing to on-vehicle cameras, mitigating privacy concerns of BS-mounted sensors. Our results demonstrate that camera-guided adaptive beamforming is practical for resilient, real-time V2X connectivity. Broad adoption of double-directional links from channel access can further increase cell size, reduce BS density, and lower handover frequency.

\section*{Acknowledgment}
This work was supported in part by the National Science Foundation (NSF) under Grants 2030141, 2030272, and 2112471. The authors would also like to thank E. Biswas and S. Shin for their assistance with conducting the outdoor experiments.
\bibliographystyle{IEEEtran}
\bibliography{standardized_mmwave_refs}


\end{document}